\documentclass[12pt]{report}
\begin{document}

\begin{center}{\bf\LARGE Solving particle-antiparticle and cosmological constant problems}\end{center}
\begin{center}Felix M Lev\end{center}
\begin{center}Email: felixlev314@gmail.com\end{center}
\begin{abstract}
We solve the particle-antiparticle and cosmological constant problems proceeding from quantum theory, which postulates that:
 various states of the system under consideration are elements of a Hilbert space
$\cal{H}$ with a positive definite metric; each physical quantity is defined by a self-adjoint operator in $\cal{H}$;
symmetry at the quantum level is defined by a representation of a real Lie algebra $A$ in $\cal{H}$ such that the representation operator of any basis element of 
$A$ is self-adjoint. 
These conditions guarantee the probabilistic interpretation of quantum theory. We explain that in the approaches to solving these problems that are described in the literature, not all of these conditions have been met.
We argue that fundamental objects in particle theory are not elementary particles and antiparticles but objects described by irreducible representations (IRs) of the de Sitter (dS) algebra.
One might ask why, then, experimental data give the impression that particles and antiparticles are fundamental and there are conserved additive quantum numbers
(electric charge, baryon quantum number and others). The matter is that, at the present stage of the universe, the contraction parameter $R$ from the dS to the Poincare algebra is very large and, in the formal limit $R\to\infty$,
one IR of the dS algebra splits into two IRs of the Poincare algebra corresponding to a particle and
its antiparticle with the same masses. The problem why the quantities $(c,\hbar,R)$ are as are does not arise because they are contraction parameters for transitions from more general Lie algebras to less general ones. Then the baryon asymmetry of the universe
problem does not arise. At the present stage of the universe, the phenomenon of cosmological acceleration (PCA)  is described without uncertainties as an inevitable {\it kinematical} consequence of quantum theory in semiclassical approximation. In particular, it is not necessary to involve dark energy the physical meaning of
which is a mystery. In our approach, background space and its geometry are not used and $R$ has nothing to do with the radius of dS space. In semiclassical approximation, the results for PCA are the same as in General Relativity if 
$\Lambda=3/R^2$, i.e., $\Lambda>0$ and there is
no freedom in choosing the value of $\Lambda$. 
\end{abstract}
\begin{flushleft}Keywords: irreducible representations; particle-antiparticle; cosmological acceleration; baryon asymmetry of the universe \end{flushleft}

\tableofcontents

\chapter{General principles of quantum theory}

In this paper we solve the particle-antiparticle and cosmological constant problems proceeding from quantum theory, which postulates:

$\cal{H}$) Various states of the system under consideration are elements of a Hilbert space
$\cal{H}$ with a positive definite metric, that is, the norm of any non-zero element of $\cal{H}$ is positive. 

$\cal{O}$) Each physical quantity is defined by a self-adjoint operator $\cal{O}$ in $\cal{H}$.

$\cal{S}$) Symmetry at the quantum level is defined by a self-adjoint representation of a real Lie algebra $A$ in $\cal{H}$ such that the representation operator of any basis element of 
$A$ is self-adjoint.

These conditions guarantee the probabilistic interpretation of quantum theory. We explain below that in the approaches to solving these problems that are described in the literature, not all of these conditions have been met.

\section{Problems with space-time background in quantum theory}
\label{DirEq}

Modern fundamental particle theories (QED, QCD and electroweak theory) are based on 
the concept of particle-antiparticle. Historically, this concept has arisen as a consequence
of the fact that the Dirac equation has solutions with positive and negative
energies. The solutions with positive energies are associated with particles, and the
solutions with negative energies --- with corresponding antiparticles. And when the positron
was found, it was a great success of the Dirac equation. Another
great success is that in the approximation $(v/c)^2$, the Dirac equation reproduces the fine structure of the hydrogen atom with a very high accuracy. 

However, now we know that there are problems with the physical interpretation of the Dirac equation. For example, in higher order approximations,
the probabilistic interpretation of non-quantized Dirac spinors is lost because they are described by representations induced from non-self-adjoined representations of the Lorenz algebra. Moreover, this problem exists for any functions described by local relativistic covariant equations (Klein-Gordon,
Dirac, Rarita-Schwinger and others). {\it So, a space of functions satisfying a local
covariant equation does not satisfy the conditions} ($\cal{H,O,S}$).

As shown by Pauli \cite{Pauli},
in the case of fields with an integer spin it is not possible to define a positive-definite charge operator while in the case of
fields with a half-integer spin it is not possible to define a positive-definite energy operator.

Another fundamental problem in the interpretation of the Dirac equation is as follows.
One of the key principles of quantum theory is the principle of superposition. This principle
states that if $\psi_1$ and $\psi_2$ are possible states of a physical system then
$c_1\psi_1+c_2\psi_2$, when $c_1$ and $c_2$ are complex coefficients, also is a
possible state. The Dirac equation is the linear equation, and, if $\psi_1(x)$ and
$\psi_2(x)$ are solutions of the equation, then $c_1\psi_1(x)+c_2\psi_2(x)$
also is a solution. In the spirit of the
Dirac equation, there should be no separate particles the electron and the positron.
It should be only one particle such that 
electron states are the states of this particle with positive energies, positron states
are the states of this particle with negative energies and the superposition of
electron and positron states should not be prohibited. However,
in view of charge conservation, baryon number conservation 
and lepton numbers conservation, the superposition of a particle and its antiparticle is prohibited. 

\begin{sloppypar}
Modern particle theories are based on Poincare symmetry which, according to ${\cal S}$), is defined by a self-adjoint representation of the Poincare algebra.  In these theories,
elementary particles, by definition, are described by self-adjoined irreducible representations (IRs) of the
Poincare algebra. Such IRs have a property that energies in them can be either 
strictly positive or strictly negative but
there are no IRs where energies have different signs. The objects described by positive-energy IRs are called particles, the objects described by negative-energy IRs are called antiparticles, 
and their energies become positive after second quantization.
There are no elementary particles which are superpositions of
a particle and its antiparticle, and as noted above, this is not in the spirit
of the Dirac equation.
\end{sloppypar}

The problems in interpreting non-quantized solutions of the Dirac equation are well known, but they are described to illustrate the problems that arise when trying to describe a particle and its antiparticle within the framework of solutions of a non-quantized local covariant equation.

In particle theories, only quantized Dirac spinors $\psi(x)$ are used. 
However, there are also problems in interpreting quantized solutions of the Dirac equation.
Here $x$ is treated as a point in Minkowski space.
However, $\psi(x)$ is an operator in the Fock space for an infinite number of particles. Each particle in the Fock space can be described by its own coordinates in the approximation when the position operator exists \cite{book}.
{\it Then the following question arises: why do we need an extra coordinate $x$ which does not have any physical meaning because it does not
belong to any particle and so is not measurable?} If we accept that physical quantities should be treated in the framework of ${\cal O}$) then $x$ is not a physical quantity because
there is no self-adjoint operator for $x$. 

A justification of the presence of $x$ in quantized solutions of local covariant equations is that in quantum field theories (QFT) the Lagrangian density 
depends on $x$, but this is only the integration parameter in the intermediate stage. The goal of
the theory is to construct the S-matrix, and, when the theory is already constructed, one can forget about
Minkowski space because no physical quantity depends on $x$. This is in the spirit of the Heisenberg
S-matrix program according to which in relativistic quantum theory it is possible to describe only
transitions of states from the infinite past when $t\to -\infty$ to the distant future when 
$t\to \infty$.

The fact that the theory gives the S-matrix in momentum representation does not
mean that the coordinate description is excluded. In typical situations, the position operator
in momentum representation exists not only in the nonrelativistic case but in the relativistic case
as well. It is known as the Newton-Wigner position operator 
\cite{NW} or its modifications. However, the
coordinate description of elementary particles can work only in some approximations. In particular, even in most favorable scenarios, for a massive particle with the mass $m$, 
its coordinates cannot be
measured with the accuracy better than the particle Compton wave length $\hbar/mc$.

When there are many bodies, the impression may arise that they are in some
space but this is only an impression. Background space-time (e.g., Minkowski space) is only a mathematical concept needed in classical theory. For example, in QED we deal with electrons, positrons and photons. 
When the position operator exists, each particle can be described by its own coordinates. In quantum theory the coordinates of Minkowski space do not have a physical meaning because they are not described by self-adjoined operators, do not refer to any particle and are not measurable. However, in classical
electrodynamics we do not consider electrons, positrons and photons. Here the concepts
of the electric and magnetic fields $({\bf E}(x),{\bf B}(x))$ have the meaning of
the mean contribution of all particles in the point $x$ of Minkowski space.

This situation is analogous to that in statistical physics. Here we do not consider each particle
separately but describe the mean contribution of all particles by temperature, pressure etc.
Those quantities have a physical meaning not for each separate particle but for ensembles
of many particles. 

Space-time background is the basic element of QFT. There is no branch of 
science where so impressive agreements between theory and experiment have been
achieved. However, those successes have been achieved only in perturbation theory
while it is not known how the theory works beyond that theory. Also,
the level of 
mathematical rigor in QFT is very poor and, as a result, QFT has several known 
difficulties and inconsistencies. 

One of the key inconsistencies of QFT is the following. 
It is known (see e.g., the textbook \cite{Bogolubov}) that quantum interacting local fields
can be treated only as operatorial distributions. A known fact from the theory of 
distributions is that the product of distributions at the same point is not a correct mathematical
operation. Physicists often ignore this problem and
use such products because, in their opinion, it preserves locality (although the operator of  $x$ does not exist). As a consequence, the representation operators of interacting systems 
in QFT are not well defined and the theory contains anomalies and divergences. 
While in renormalizable
theories the problem of divergences can be circumvented at the level of perturbation
theory, in quantum gravity divergences cannot be excluded even in lowest orders of
perturbation theory. As noted above, in spite of such mathematical problems, QFT is very popular since it has achieved successes in describing many experimental data.

In the present paper, we consider particle-antiparticle and cosmological constant problems.
In our approach, for solving those problems there is no need to involve space-time background and the problems can be solved using only rigorous mathematics.

\section{Symmetry at quantum level}
\label{symmetry} 

In the literature, symmetry in QFT is usually explained as follows.
Since Poincare group is the group of motions of Minkowski space, the system under consideration should be described by unitary representations of this group. 
This implies that the representation generators commute according to the commutation relations of the Poincare group Lie algebra:
\begin{eqnarray}
&[P^{\mu},P^{\nu}]=0,\quad [P^{\mu},M^{\nu\rho}]=-i(\eta^{\mu\rho}P^{\nu}-
\eta^{\mu\nu}P^{\rho}),\nonumber\\
&[M^{\mu\nu},M^{\rho\sigma}]=-i (\eta^{\mu\rho}M^{\nu\sigma}+\eta^{\nu\sigma}M^{\mu\rho}-
\eta^{\mu\sigma}M^{\nu\rho}-\eta^{\nu\rho}M^{\mu\sigma})
\label{PCR}
\end{eqnarray}
where $\mu,\nu=0,1,2,3$, $\eta^{\mu\nu}=0$ if $\mu\neq \nu$, $\eta^{00}=-\eta^{11}=
-\eta^{22}=-\eta^{33}=1$,
$P^{\mu}$ are the operators of the four-momentum and  $M^{\mu\nu}$ are the operators of Lorentz angular momenta. This approach is in the spirit of 
the Erlangen Program proposed by Felix Klein in 1872 when quantum theory did not yet exist.
However, although the Poincare group is the group of motions of Minkowski space, 
the description (\ref{PCR}) does not involve this group and this space.

As noted in Sec. \ref{DirEq}, background space is only a mathematical concept: in quantum theory, each physical quantity should be described by an operator but there are no operators for the coordinates of background space. {\it There is no law that every physical theory must contain a background space.} For example, it is not used in nonrelativistic quantum mechanics and in IRs describing elementary particles. In particle theory, transformations from the
Poincare group are not used because, according to the Heisenberg $S$-matrix program, it is possible to describe only transitions of states from the infinite past when $t\to -\infty$ to the distant future 
when $t\to +\infty$.  In this theory, systems are described by observable physical quantities --- momenta and angular momenta. So, {\it symmetry at the quantum level is defined not by a background space and its group of motions but by the condition ${\cal S}$)} (see \cite{book, DS} for more details). In particular,
Eq. (\ref{PCR}) can be treated {\it as the definition of relativistic invariance at the
quantum level.} 

Then each elementary particle is described
by a self-adjoined IR of a real Lie algebra $A$ and a system of $N$ noninteracting particles is
described by the tensor product of the corresponding IRs. This implies that, for the system as a whole, each momentum operator  is a sum of the corresponding single-particle momenta, 
each angular momentum operator is a sum of the corresponding single-particle angular momenta, and {\it this is the most complete possible description of this system}. 
In particular, nonrelativistic symmetry implies that $A$ is the Galilei algebra, relativistic symmetry implies that $A$ is the Poincare algebra, de Sitter (dS) symmetry implies that $A$ is the dS algebra so(1,4) and anti-de Sitter (AdS) symmetry implies that $A$ is the AdS algebra so(2,3).

In his famous paper "Missed Opportunities" \cite{Dyson} Dyson notes that: 
\begin{itemize}
\item a) Relativistic
quantum theories are more general than nonrelativistic quantum theories
even from purely mathematical considerations because Poincare group is more symmetric
than Galilei one: the latter can be obtained from the former by contraction $c\to\infty$. 
\item b) dS and AdS 
quantum theories are more general than relativistic quantum theories
even from purely mathematical considerations because dS and AdS groups are more symmetric
than Poincare one: the latter can be obtained from the former by contraction $R\to\infty$
where $R$ is a parameter with the dimension $length$, and the meaning of this parameter
will be explained below.   
\item c) At the same time, since dS and AdS groups are semisimple, they have a maximum possible symmetry and cannot be obtained from more symmetric groups by contraction. 
\end{itemize}

As noted above, symmetry at the quantum level should be defined in the framework 
of ${\cal S}$), and in \cite{book}, the
statements a)-c) have been reformulated in terms of the corresponding Lie algebras. It has
also been shown that the fact that quantum theory is more general than classical theory follows even from purely mathematical considerations
because formally the classical symmetry algebra can be obtained from the symmetry algebra
in quantum theory by contraction $\hbar\to 0$. For these reasons, the most general description in terms of ten-dimensional Lie algebras should be carried out in terms of quantum dS or AdS symmetry. However, as explained below, in particle theory, dS symmetry is more general than
AdS one.

The definition of those symmetries is as follows. 
If $M^{ab}$ ($a,b=0,1,2,3,4$, $M^{ab}=-M^{ba}$) are the angular momentum operators for the system under consideration, they should satisfy the
commutation relations:
\begin{equation}
[M^{ab},M^{cd}]=-i (\eta^{ac}M^{bd}+\eta^{bd}M^{ac}-
\eta^{ad}M^{bc}-\eta^{bc}M^{ad})
\label{CR}
\end{equation}
where $\eta^{ab}=0$ if $a\neq b$, 
$\eta^{00}=-\eta^{11}=-\eta^{22}=-\eta^{33}=1$ and 
$\eta^{44}=\mp 1$ for the dS and AdS symmetries, respectively.

Although the dS and AdS groups are the groups of motions of dS and AdS spaces, respectively,
the description in terms of (\ref{CR}) does not involve those groups and spaces,
and {\it it is a definition of dS and AdS symmetries in the framework of ${\cal S}$)} (see the discussion in \cite{book,DS}). 
In QFT, interacting particles are described by field functions defined on Minkowski, 
dS and AdS spaces. However, since we consider
only noninteracting bodies and describe them in terms of IRs, at this level we don't need these fields and spaces.

The procedure of contraction from dS or AdS symmetry to Poincare one is defined as follows. If we {\it define} the momentum
operators $P^{\mu}$ as $P^{\mu}=M^{4\mu}/R$ ($\mu=0,1,2,3$) then in the formal
limit when $R\to\infty$, $M^{4\mu}\to\infty$ but the quantities
$P^{\mu}$ are finite, Eq. (\ref{CR}) become Eq. (\ref{PCR}). Here {\it R is a parameter which has nothing to do with the dS and AdS spaces}. 
As seen from Eq. (\ref{CR}), quantum dS and AdS theories
do not involve the dimensional parameters $(c,\hbar,R)$ because $(kg,m,s)$ are meaningful only at the macroscopic level. 

As noted by Berry \cite{Berry}, the reduction from more general theories to less general ones 
involves a quantity $\delta$ which is not equal to zero in more general theories and becomes zero
in less general theories. This reduction involves the study of limits and is often obstructed by the fact that the limit is singular. In 
\cite{Berry}, several examples of such reductions are considered. However, at the quantum level, the reduction (contraction) should be described in terms of relations between the representation operations of more general and less general algebras.
As explained in \cite{book}, in the limit when the contraction parameter goes to zero or infinity, some original
representation operators become singular (in agreement with the results of \cite{Berry}). However, 
it is possible to define a new set of operators such that they remain finite in this limit. Then, in less
general theories, some commutators become zero while in more general theories they are non-zero.
So, less general theories contain more zero commutators then corresponding more general theories. 

Probably, the most known case is the
reduction from relativistic to nonrelativistic theory.
In relativistic theory, the quantity $c$ is not needed, velocities ${\bf v}$ are dimensionless and, if $v=|{\bf v}|$
then $v\leq 1$ if tachyons are not taken into account. However, if people want to describe velocities in $m/s$
then $c$ also has the dimension $m/s$. Physicists usually understand that physics cannot (and should not) derive that $c\approx 3 \cdot 10^8m/s$. This value is purely kinematical (i.e., it does not depend on gravity and other interactions) and is as is simply because people
want to describe velocities in $m/s$. Since the quantities ($m,s$) have a physical meaning only at the macroscopic level, one can expect that the values of $c$ in $m/s$ are different at different
stages of the universe. In \cite{Berry}, the connection between relativistic and 
nonrelativistic theories is described in the ''low-speed'' series expansions in $\delta=v/c$. 
However, such expansions are well defined only in classical (non-quantum) theory.
At the quantum level, this reduction should be described in terms of relations between the representation operations of the Poincare and Galilei algebras. Then, in agreement with  \cite{Berry}, the transition
from relativistic to nonrelativistic theory becomes singular in the formal limit $c\to\infty$. 
As described in \cite{book,Bargmann}, the singularities can be resolved by 
using the Galilei boost operators $G^j=M^{0j}/c$, $(j=1,2,3)$
instead of the Poincare boost operators $M^{0j}$ and by using the time translation operator
$E=P^0c$ instead of the Poincare energy operator $P^0$.  Then, as follows from Eq. (\ref{PCR}), 
instead of the relations $[M^{0j},M^{0k}]=-iM^{jk}$ where $j,k=1,2,3,\,\, j\neq k$, we have
$[G^j,G^k]=-iM^{jk}/c^2$.

So far, no approximations have been made. A question arises whether the strong limits of the
operators $M^{jk}/c^2$ are zero when $c\to\infty$. In general, not for all elements $x$ of the Hilbert space
under consideration, $y=(M^{jk}/c^2)x$ become zero when $c\to\infty$. The meaning
of the nonrelativistic approximation at the operator level is that only those elements $x$
are important for which $y\to 0$ when $c\to\infty$. Therefore, in the nonrelativistic approximation,
$[G^j,G^k]=0$ and we have a greater number of zero commutators because in the
relativistic case, $[M^{0j},M^{0k}]\neq 0$. And, since $M^{0j}=G^jc$, we conclude
that, when $c\to\infty$, the operators $M^{0j}$ become singular in agreement with the observation
in \cite{Berry}.

Consider now the relation between classical and quantum theories. In the latter, the quantity $\hbar$ is not needed and angular momenta are dimensionless. As shown even in textbooks, their projections
can take only the values multiple to $\pm 1/2$. However, when people want to describe angular momenta in 
$kg\cdot m^2/s$, $\hbar$ and all the operators in Eq. (\ref{CR}) become dimensional and also have the dimension $kg\cdot m^2/s$. Then all 
nonzero commutators in the symmetry algebra become proportional to $\hbar$ and Eq. (\ref{CR}) can
be represented as $[M_{ab},M_{cd}]=i\hbar A_{abcd}$.

Physicists usually understand that physics cannot (and should not) derive that $\hbar \approx 1.054 \cdot 10^{-34}kg\cdot m^2/s$. This value is purely kinematical and is as is simply because people
want to describe angular momenta in $kg\cdot m^2/s$.  Since the quantities ($kg,m,s$) have a physical meaning only at the macroscopic level, one can expect that the values of $\hbar$ in $kg\cdot m^2/s$ are different at different stages of the universe.  If $A_{abcd}\neq 0$ then, in general, 
not for all elements $x$ of the Hilbert space
under consideration, $y=\hbar A_{abcd}x$ become zero when $\hbar\to 0$. The meaning
of the classical approximation is that only those elements $x$
are important for which $y\to 0$ when $\hbar\to 0$. Therefore, in this approximation,
all the commutators become zero and all physical quantities are defined without
uncertainties. So, even the description in terms of Hilbert spaces becomes redundant.

Typically, in particle theories, the quantities
$c$ and $\hbar$ are not involved and it is said that the units $c=\hbar=1$ are used.

At the quantum level, Eq. (\ref{CR}) is the most general description
of dS and AdS symmetries and all the operators in Eq. (\ref{CR}) are dimensionless. At this level,
the theory does not need the quantity $R$ and, in full analogy with the above discussion of the quantities
$c$ and $\hbar$, one can say $R=1$ is a possible choice. The dimensional quantity $R$ arises if, instead of the dimensionless operators
$M^{4\mu}$, physicists want to deal with the 4-momenta $P^{\mu}$ {\it defined} such that
$M^{4\mu}=RP^{\mu}$. In full analogy with the discussion of $c$ and $\hbar$, physics 
cannot (and should not) derive the value of $R$. It
is as is simply because people want to measure distances in meters. This value is purely kinematical, i.e., it does not depend on gravity and other interactions. As noted in Sec. \ref{discussion},
at the present stage of the universe, $R$ is of the order of $10^{26}m$ but, since the concept of
meter has a physical meaning only at the macroscopic level, one can expect that the values of $R$ in 
meters are different at different stages of the universe.

Although, at the level of contraction parameters, $R$ has nothing to do with the radius of the background space and is fundamental to the same extent as $c$ and
$\hbar$, physicists usually want to
treat $R$ as the radius of the background space. In General Relativity (GR) which is the
non-quantum theory, the cosmological constant $\Lambda$ equals $\pm 3/R^2$ for
the dS and AdS symmetries, respectively. Physicists usually believe that physics should derive the value 
of $\Lambda$ and that the solution to the dark energy problem depends on this value. 
They also believe that QFT of gravity should confirm the experimental result that, in units
$c=\hbar=1$, $\Lambda$ is of the order
of $10^{-122}/G$ where $G$ is the gravitational constant. We will discuss this
problem in Sec. \ref{discussion}.

As follows from Eq. (\ref{CR}), $[M^{4\mu},M^{4\nu}]=iM^{\mu\nu}$. Therefore 
$[P^{\mu},P^{\nu}]=iM^{\mu\nu}/R^2$. A question arises whether the strong limits of the
operators $M^{\mu\nu}/R^2$ are zero when $R\to\infty$. In general, not for all elements $x$ of the Hilbert space
under consideration, $y=(M^{\mu\nu}/R^2)x$ become zero when $R\to\infty$. The meaning
of the Poincare approximation at the operator level is that only those elements $x$
are important for which $y\to 0$ when $R\to\infty$. Therefore, in the Poincare approximation,
$[P^{\mu},P^{\nu}]=0$ and we have a greater number of zero commutators because in the
dS and AdS cases, $[M^{4\mu},M^{4\nu}]\neq 0$. And, since $M^{4\mu}=P^{\mu}R$, we conclude
that, when $R\to\infty$, the operators $M^{4\mu}$ become singular in agreement with the observation
in \cite{Berry}.

\chapter{Solving particle-antiparticle problem}

\section{Particles and antiparticles in standard quantum theory}
\label{PoincareIRs}

Standard particle theories are based on Poincare symmetry, and here the
concepts of particles and antiparticles are considered from the point of view of
two approaches which we call Approach A and Approach B. We first recall the basic known facts about IRs of the
Poincare algebra. Their classification has been first given by Wigner \cite{Wigner} and then repeated by many authors (see e.g., \cite{Novozh}).

We denote $E=P^0$ the energy operator and ${\bf P}=(P^1,P^2,P^3)$ the spatial momentum
operator. Then $W=E^2-{\bf P}^2$ is the Casimir operator of the Poincare algebra, i.e., it
commutes with all operators of the algebra. As follows from the Schur lemma, $W$ has only one eigenvalue in every IR. We will not consider tachyons and then this
eigenvalue is $\geq 0$ and can be denoted $m^2$ where $m\geq 0$ is called the particle mass.
We will consider massive IRs where $m>0$ and the case $m=0$ will be mentioned below.

Let $p$ be the particle four-momentum such that $p^2=(p^0)^2-{\bf p}^2=m^2$. We denote
$v=p/m$ the particle four-velocity such that $v^2=1$. Then $v_0^2=1+{\bf v}^2$ and
we will always choose $v_0$ such that $v_0\geq 1$. Let 
$d\rho({\bf v})=d^3{\bf v}/v_0$
be Lorentz invariant volume element on the Lorentz hyperboloid. 
If $s$ is the spin of the particle
under consideration, then we use $||...||$ to denote the norm
in the space of unitary IR of the group SU(2) with the spin $s$. Then
the space of a self-adjoned IR of the Poincare algebra is the space of functions 
$f({\bf v})$ on the Lorentz hyperboloid with the range
in the space of IR of the group SU(2) with the spin $s$ and
such that
$$ \int\nolimits ||f({\bf v})||^2\rho({\bf v}) <\infty$$
Then the operators of the IR are given by \cite{Wigner,Novozh}
\begin{equation}
{\bf J}=l({\bf v})+{\bf s},\quad {\bf N}=-i v_0
\frac{\partial}{\partial {\bf v}}+\frac{{\bf s}\times {\bf v}}
{v_0+1} ,\quad P=\pm mv
\label{PIR}
\end{equation}
where ${\bf J}=\{M^{23},M^{31},M^{12}\}$, ${\bf
N}=\{M^{01},M^{02},M^{03}\}$, ${\bf s}$ is the spin operator,
${\bf l}({\bf v})=-i{\bf v}\times \partial/\partial {\bf
v}$ and $\pm$ refers to the IRs with positive and negative energies, respectively.

Approach A is based on the fact that, as follows from Eq. (\ref{PIR}), in self-adjoined IRs of the
Poincare algebra, the energy spectrum can be either $\geq 0$ or $\leq 0$,
and there are no IRs where the energy spectrum contains both, positive and
negative energies. In this approach, the objects described by the corresponding IRs are called
elementary particles and antiparticles, respectively. 
On the other hand, Approach B proceeds from the assumptions that elementary particles
are described by local covariant equations. The solutions of these equations with positive
energies are called particles and solutions with negative energies are called antiparticles

When we consider a system consisting of particles and antiparticles, the energy signs for both of them should be the same. Indeed, consider, for example a system of two particles with the same mass, 
and let their momenta ${\bf p}_1$ and ${\bf p}_2$ be such that ${\bf p}_1+{\bf p}_2=0$. Then, if the energy of particle 1 is positive, and the energy of particle 2 is negative then the total four-momentum of the system would be zero what contradicts experimental data. By convention, the energy sign of all  particles and antiparticles
in question is chosen to be positive. For this purpose, the procedure of second quantization is defined such that
after this procedure the energies of antiparticles become positive. Then the mass of any particle is the minimum value of its energy. 

Suppose now that we have two particles such that particle 1 has the mass $m_1$, spin $s_1$ and is characterized by some additive quantum numbers (e.g., electric charge, baryon quantum 
number etc.), and particle 2 has the mass $m_2$, spin $s_2=s_1$ and all additive quantum numbers
characterizing particle 2 equal the corresponding quantum numbers for particle 1 with the opposite sign.
A question arises when particle 2 can be treated as an antiparticle for particle 1. Is it necessary that $m_1$ 
should be exactly equal $m_2$ or $m_1$ and $m_2$ can slightly differ each other? In particular, can we guarantee that the mass of the positron exactly equals the mass of the electron, the mass of the proton
exactly equals the mass of the antiproton etc.? 
If we work only in the framework
of Approach A then we cannot answer this question because here IRs for particles 1 and 2 are independent on each other and there are no
limitations on the relation between $m_1$ and $m_2$. 

On the other hand, in Approach B, $m_1=m_2$ but this has been achieved
at the expense of losing probabilistic interpretation. Indeed, here, a particle and its antiparticle are elements of the same field state $\psi (x)$ with positive
and negative energies, respectively, where $x$ is a vector from Minkowski space
and $\psi (x)$ satisfies a relativistic covariant field equation (Dirac, Klein-Gordon, Rarita-Schwinger and others). However, it has been already noted in Sec. \ref{DirEq} that, 
at the  quantum level, covariant fields and the quantity $x$ are not defined in the
framework of (${\cal H,O,S}$). In particular, at the quantum level, the physical meaning of $x$ is unclear because there is no operator for $x$. 

A usual phrase in the literature is that in QFT, the fact that $m_1=m_2$ follows from the CPT theorem. As shown e.g., in \cite{Luders,CPT}, it is a consequence of locality since, {\it by construction}, states
described by local covariant equations are direct sums of IRs for a particle and
its antiparticle with equal masses. However, since the concept of locality is not formulated in 
the framework of (${\cal H,O,S}$), this concept does not have a clear physical meaning, and this fact has been pointed out even in known textbooks (see e.g., \cite{Bogolubov}). Therefore, QFT does not give a rigorous proof that $m_1=m_2$.

Also, can one pose the question what is happening if locality is only an approximation: in that case the equality of masses is exact or approximate? 
However, since, at the quantum level, the physical meaning of the concept of locality is unclear, the physical meaning of this question is also unclear. Consider a simple model when electromagnetic and weak interactions are absent. Then the fact that the proton and the neutron have equal masses has nothing to do with locality; it is only a consequence of the fact that they belong to the same isotopic multiplet, i.e., they are simply different states of the same object---the nucleon. 

Note that in Poincare invariant quantum theories, there
can exist elementary particles for which all additive quantum numbers are zero. Such particles are called neutral because they coincide with their antiparticles.

\section{Particles and antiparticles in AdS quantum theories}
\label{subsecAdS}

In theories where the symmetry algebra is the AdS algebra, the structure of IRs
is known (see e.g., \cite{book,Evans}). The operator $M^{04}$ is the AdS analog of the energy
operator. Let $W$ be
the Casimir operator $W=\frac{1}{2}\sum M^{ab}M_{ab}$ where a sum over repeated indices is assumed. Here lowering and raising indices are carried out using the tensor $\eta^{ab}$ defined in Sec. \ref{symmetry} and, as noted after Eq. (\ref{CR}), $\eta^{44}=1$ for the AdS case.
As follows from the Schur lemma, the operator $W$ has only one eigenvalue in every IR. By analogy with
Poincare quantum theory, we will not consider AdS tachyons and then one can define the AdS 
mass $\mu$ such that $\mu\geq 0$ and $\mu^2$ is the
eigenvalue of the operator $W$. 

As noted in Sec. \ref{symmetry}, the procedure of contraction from the AdS algebra to the Poincare one is defined in terms of the parameter $R$ such that $M^{\nu 4}=RP^{\nu}$.
This procedure has a physical meaning only if $R$ is rather large. In that case
the AdS mass $\mu$ and the Poincare mass $m$ are related as $\mu=Rm$, and the relation between the
AdS and Poincare energies is analogous. Since AdS symmetry is more general then Poincare one then 
$\mu$ is more general than $m$. In contrast to the Poincare masses and energies, the AdS masses
and energies are dimensionless. As noted in Sec. \ref{discussion}, at the present stage
of the universe $R$ is of the order of $10^{26}m$. Then the AdS masses of the electron,
the Earth and the Sun are of the order of $10^{39}$, $10^{93}$ and $10^{99}$, respectively. 
The fact that even the AdS mass of the electron is so large might be an
indication that the electron is not a true elementary particle.
In addition, the present upper level for the photon mass is $10^{-17}ev$. This value
seems to be an extremely tiny quantity. However, the corresponding AdS mass is of the order of $10^{16}$,
and so, even the mass which is treated as extremely small in Poincare
invariant theory might be very large in AdS invariant theory.

In the AdS case, there are IRs with positive and negative energies, and they belong to the discrete series 
\cite{book,Evans}. Therefore, one can define particles and antiparticles. 
  If $\mu_1$ is the AdS mass for a positive energy IR, then the energy spectrum contains the eigenvalues
$\mu_1,\mu_1+1,\mu_1+2, ...\infty$, and, if $\mu_2$ is the AdS mass for a negative energy IR, then the energy spectrum contains the eigenvalues $-\infty,...-\mu_2-2,-\mu_2-1,-\mu_2$. 

Therefore, the situation is pretty much analogous to that in Poincare invariant theories, and, without involving local AdS invariant equations 
there is no way to conclude whether the mass of a particle equals the mass of the corresponding antiparticle. These equations describe local fields in the AdS space. In view of what was said above about the background space in QFT, these fields are not defined within the framework
of (${\cal H,O,S}$). Therefore, in AdS invariant theory, just as in the case of Poincare invariant theory, within the framework
of (${\cal H,O,S}$) it is also impossible to prove that the mass of a particle equals the mass of the corresponding antiparticle.

Since Poincare quantum theory is obtained from AdS quantum theory by contraction
$R\to\infty$ and $m=\mu /R$ then Poincare massless IRs are obtained from AdS IRs
not only when $\mu=0$ but when $\mu$ is any finite number. In Poincare quantum theories,
massless particles are characterized such that for them helicity is the conserved quantum number. For this reason, as shown in \cite{Evans} (see also \cite{book}),
the AdS massless particles are described by IRs where $\mu=2+s$. Before the discovery of
neutrino oscillations, neutrinos were treated as massless with the left-handed
helicity and antineutrinos --- as massless with the right-handed helicity, but now
they are treated as massive particles. 
The photon is usually treated as massless although, as noted in \cite{Okun},
QED will not be broken if the photon has a small nonzero mass. In contrast to the neutrino case,
it is described not by IRs of the purely Poincare algebra but by IRs of the Poincare algebra
with spatial reflections added (see e.g., \cite{Novozh}). For this reason, the photon
is the neutral particle because it coincides with its own antiparticle.

\section{Problems with the definition of particles and antiparticles in dS quantum theories}
\label{subsecdS}

In this section we explain why the description of particles and antiparticles in the case of dS symmetry considerably differs from that in the cases of Poincare and AdS symmetries described in the preceding sections.

The Casimir operator $W=\frac{1}{2}\sum M^{ab}M_{ab}$ is now defined in the same way as
in the AdS case but, as noted after Eq. (\ref{CR}), $\eta^{44}=-1$ for the dS case.
By analogy with the AdS case, it follows from the Schur lemma that the operator $W$ has only one eigenvalue in every IR, one can define the dS 
mass $\mu$ such that $\mu\geq 0$ and $\mu^2$ is the
eigenvalue of the operator $W$. 

In his book \cite{Mensky} Mensky describes the construction of unitary IRs of the dS group using
the theory of induced representations (see e.g., \cite{Mackey,Jorgensen}).
In \cite{book} we describe how this construction can be used for 
constructing self-adjoined IRs of the dS algebra. Here we explicitly describe two implementations
of such a construction: when the representation space is a space of functions on two
Lorentz hyperboloids and when it is a space of functions on the three-dimensional unit sphere in the four-dimensional space. 

\begin{sloppypar}
In the first case, the space of IR is the space of functions $(f_1({\bf
v}),f_2({\bf v}))$ on two Lorentz hyperboloids with the range
in the space of unitary IR of the group SU(2) with the spin $s$ and
such that
$$ \int\nolimits [||f_1({\bf v})||^2+
||f_2({\bf v})||^2]d\rho({\bf v}) <\infty $$
where, as in Sec. \ref{PoincareIRs}, ${\bf s}$ is the spin operator and $||...||$ is the norm
in the space of unitary IR of the group SU(2) with the spin $s$. 
\end{sloppypar}

In this case, the explicit calculation \cite{book} shows that the action of representation
operators on functions with the support on the first hyperboloid is 
\begin{eqnarray}
&&{\bf J}=l({\bf v})+{\bf s},\quad {\bf N}=-i v_0
\frac{\partial}{\partial {\bf v}}+\frac{{\bf s}\times {\bf v}}
{v_0+1} \nonumber\\
&& {\bf B}=\mu {\bf v}+i [\frac{\partial}{\partial {\bf v}}+
{\bf v}({\bf v}\frac{\partial}{\partial {\bf v}})+\frac{3}{2}{\bf v}]+
\frac{{\bf s}\times {\bf v}}{v_0+1}\nonumber\\
&& {\cal E}=\mu v_0+i v_0({\bf v}
\frac{\partial}{\partial {\bf v}}+\frac{3}{2})
\label{upperhyperb}
\end{eqnarray}
where $\mu>0$ is a parameter which can be called the dS mass, ${\bf J}$, ${\bf N}$ and ${\bf l}({\bf v})$ are given by the same expressions
as in Sec. \ref{PoincareIRs}, ${\bf B}=\{M^{41},M^{42},M^{43}\}$ and ${\cal E}=M^{40}$.
At the same time, the action on functions with the support on the second hyperboloid
is given by
\begin{eqnarray}
&&{\bf J}=l({\bf v})+{\bf s},\quad {\bf N}=-i v_0
\frac{\partial}{\partial {\bf v}}+\frac{{\bf s}\times {\bf v}}
{v_0+1} \nonumber\\
&& {\bf B}=-\mu {\bf v}-i [\frac{\partial}{\partial {\bf v}}+
{\bf v}({\bf v}\frac{\partial}{\partial {\bf v}})+\frac{3}{2}{\bf v}]-
\frac{{\bf s}\times {\bf v}}{v_0+1}\nonumber\\
&& {\cal E}=-\mu v_0-i v_0({\bf v}
\frac{\partial}{\partial {\bf v}}+\frac{3}{2})
\label{lowerhyperb}
\end{eqnarray}
Note that the expressions for the action of the Lorentz algebra
operators on the first and second hyperboloids are the same and coincide
with the corresponding expressions for IRs of the Poincare
algebra in Eq. (\ref{PIR}). At the same time, the expressions for the action of
the operators $M^{4\mu}$ on the first and second hyperboloids differ by sign.

In the second case, the representation space is the space of functions on
the group SU(2). Its elements can be represented by the
points $u=({\bf u},u_4)$ of the three-dimensional sphere $S^3$
in the four-dimensional space as $u_4+i{\bf{\sigma}}{\bf u}$
where ${\bf{\sigma}}$ are the Pauli matrices and $u_4=\pm
(1-{\bf u}^2)^{1/2}$ for the upper and lower hemispheres,
respectively. Then the
Hilbert space of the IR is the space of functions $\varphi (u)$ on
$S^3$ with the range in the space of the unitary IR of the su(2)
algebra with the spin $s$ and such that
$$ \int\nolimits ||\varphi(u)||^2du <\infty $$
where $du$ is the SO(4) invariant volume element on $S^3$. The
explicit calculation \cite{book} shows  that the operators have the form
\begin{eqnarray}
&&{\bf J}=l({\bf u})+{\bf s},\quad {\bf B}=i u_4
\frac{\partial}{\partial {\bf u}}-{\bf s}, \quad {\cal E}=(\mu +3i/2)u_4+i u_4{\bf u}
\frac{\partial}{\partial {\bf u}} \nonumber\\
&& {\bf N}=-i [\frac{\partial}{\partial {\bf u}}-
{\bf u}({\bf u}\frac{\partial}{\partial {\bf u}})]
+(\mu +3i/2){\bf u}-{\bf u}\times {\bf s}+u_4{\bf s}
\label{3sphere}
\end{eqnarray}
Since Eqs. (\ref{upperhyperb}) and (\ref{lowerhyperb}) on one
hand and Eq. (\ref{3sphere}) on  the other  are
the  different implementations of  the   same
representation, there exists a unitary operator transforming
functions $f(v)$ into $\varphi (u)$ and operators
(\ref{upperhyperb}) and (\ref{lowerhyperb}) into operators (\ref{3sphere}). For example, in
the spinless case the operators (\ref{upperhyperb}) and (\ref{3sphere}) are
related to each other by a unitary transformation
\begin{equation}
\varphi (u)=exp(-i\mu lnv_0)v_0^{3/2}f(v)
\label{unitequiv}
\end{equation}
where the relation between the points of the upper hemisphere
and the first hyperboloid is ${\bf u}={\bf v}/v_0$ and $u_4=(1-{\bf
u}^2)^{1/2}$. The relation between the points of the lower
hemisphere and the second hyperboloid is ${\bf u}=-{\bf v}/v_0$ and
$u_4=-(1-{\bf u}^2)^{1/2}$.

The equator of $S^3$ where $u_4=0$ has
measure zero with respect to the upper and lower hemispheres.
For this reason one might think that it is of no interest for
describing particles in dS theory. Nevertheless, while none of the components of $u$ has the
magnitude greater than unity, the points of the equator in terms of
velocities is characterized by the condition that $|{\bf v}|$
is infinitely large and therefore standard Poincare
momentum ${\bf p}=m{\bf v}$ is infinitely large too. This poses
a question whether ${\bf p}$ always has a physical meaning.
From mathematical point of view, Eq. (\ref{3sphere}) might seem
more convenient than Eqs. (\ref{upperhyperb}) and (\ref{lowerhyperb}) since
$S^3$ is compact and there is no need to break it into the
upper and lower hemispheres. However, Eqs. (\ref{upperhyperb}) and (\ref{lowerhyperb}) 
are convenient for investigating Poincare approximation while the expressions (\ref{3sphere}) are not convenient for this purpose because the Lorentz boost
operators ${\bf N}$ in them depend on $\mu$.

Indeed, if we {\it define} 
\begin{equation}
E=P^0={\cal E}/R,\,\, {\bf P}={\bf B}/R,\,\,m=\mu/R
\label{EPm}
\end{equation}
then in the formal limit when $R\to\infty$, $\mu\to\infty$ but $E,\,\,{\bf P}$ and $m$ remain finite,  
Eqs. (\ref{upperhyperb}) and (\ref{lowerhyperb}) become Eq. (\ref{PIR}) for positive
and negative energy IRs of the Poincare algebra, respectively. 
Therefore, dS symmetry is broken in the formal limit $R\to\infty$ because one IR of the dS algebra splits into two IRs of the Poincare algebra with positive and negative energies and with equal masses. 

Since the number of states in dS IRs is twice as big as the number of states in IRs of the Poincare
algebra, one might think that each IR describes a particle and its antiparticle simultaneously. 
But this is not true even from the fact that when we talk about a particle and its antiparticle, we mean that there are two different IRs, but in this case there is only one IR.
In addition, the question of what is the mass difference between a particle and its antiparticle if $R$ is finite has no physical meaning because, according to the Schur lemma, the operator $W$ has only one eigenvalue in this IR 
and all states have the same mass $\mu$. Another argument that this is not true is as follows.

Let us call states with the support of their wave functions on the 
first hyperboloid or on the northern hemisphere as particles and states with the support on the second hyperboloid or on the southern hemisphere as their antiparticles. {\it The physical meaning
of such definitions is problematic since there is no guaranty that
the energy of particles is always positive and the energy of antiparticles is always negative.}
Nevertheless, even with such a definition,  
states which are superpositions of a particle and its antiparticle obviously belong to the representation space under consideration, 
i.e., they are not prohibited. However, this contradicts the
superselection rule that the wave function cannot be a superposition of states with opposite electric charges, baryon and lepton quantum numbers etc. 
Therefore, in the dS case, there are no superselection rules which prohibit superpositions of states
with opposite electric charges, baryon quantum numbers etc. In addition, in this case it is not possible to
define the concept of neutral particles, i.e., particles which coincide with their antiparticles
(e.g., the photon). This question will be discussed in Chap. \ref{open}.

As noted in Sec. \ref{symmetry} and shown in the discussion of Eq. (\ref{EPm}), dS symmetry is more general than Poincare one, and the latter can be treated as
a special degenerate case of the former in the formal limit $R\to\infty$. This means that, with any desired accuracy,
any phenomenon described in the framework of Poincare symmetry can be also described in the framework of
dS symmetry if $R$ is chosen to be sufficiently large, but there also exist phenomena for explanation of which
it is important that $R$ is finite and not infinitely large (see \cite{book}). 

The fact that dS
symmetry is higher than Poincare one is clear even from the fact that, in the framework of the latter symmetry,
it is not possible to describe states which are superpositions of states on the upper and lower hemispheres.
Therefore, breaking one dS IR into two independent IRs defined on the northern and southern hemispheres
obviously breaks the initial symmetry of the problem.  This fact is in agreement
with the Dyson observation (mentioned above) that dS group is more symmetric than Poincare one.
 
When $R\to\infty$, standard concepts of particle-antiparticle, electric charge and baryon 
and lepton quantum numbers are restored, i.e., in this limit superpositions of particle and antiparticle states become prohibited according to the superselection rules.  {\it Therefore, those concepts have a rigorous physical meaning only as a result of symmetry breaking at $R\to\infty$, but if $R$ is finite they can be only good approximations when $R$ is rather large.}

The observable equality of masses of particles and their corresponding antiparticles can be now explained as a consequence of the fact that observable properties of elementary particles can be described not by exact Poincare symmetry but by dS symmetry with a very large but finite value of $R$. In this approximation,
for combining a particle and its antiparticle into one object, there in no need to assume
locality and involve local field functions because
 a particle and its antiparticle already belong to the same IR
of the dS algebra (compare with the above remark about the isotopic symmetry in the proton-neutron system). As noted above, in this approximation it is not correct to pose the question
about the mass difference between a particle and its antiparticle because they have the same mass $\mu$. However, it is correct to pose the following problem.

When $R$ is finite but very large, the concepts of electric charge and baryon number are not precise, but make sense with very high accuracy. Let us assume that our experiment shows that there are particles with electric charge $e$ and $-e$. In the formal limit $R\to\infty$
there can be no particles which are superpositions of states with the charges $e$ and $-e$.
However, in the approximation when $R$ is very large but finite and the concept of electric charge is meaningful with very high accuracy, such superpositions are possible. In that case,
if in some experiment we observe protons then with a very small probability we can observe
antiprotons. For example, if in some experiment we observe elastic scattering of protons
on a neutral target $T$, $p+T\to  p+T$ then with a very small probability we will observe the process $p+T\to {\bar p}+T$. With the current value of $R$, the probability of such a process is negligible, but in the early stages of the universe it can be noticeable. 
But the calculation of the probability of such a process can only be carried out when a particle theory based on dS symmetry rather than Poincare symmetry, is constructed.

\begin{sloppypar}
\section{dS vs. AdS and baryon asymmetry of the universe problem}
\label{asymm}
\end{sloppypar}

In this chapter we have discussed how the concepts of particles and antiparticles should be defined in the cases of Poincare, AdS and dS symmetries.  In the first two cases, the situations are similar: IRs where the energies are $\geq 0$ are
treated as particles, and IRs where the energies are $\leq 0$ are
treated as antiparticles. Then a problem arises how to prove that the masses of a particle and
the corresponding antiparticle are the same. As noted in Secs. \ref{PoincareIRs} and 
\ref{subsecAdS}, without involving local covariant equations 
there is no way to conclude whether it is the case. Since the concept of locality is not formulated  
in the framework of (${\cal H,O,S}$), QFT does not give a rigorous proof
that the masses of a particle and the corresponding antiparticle are the same. 

As described in Sec. \ref{subsecdS}, in the case of dS symmetry, the approach to the concept of particle-antiparticle is radically different from the approaches in the cases of Poincare and
AdS symmetries. Here, the fundamental objects are not particles and antiparticles, but objects  described by self-adjoined IRs of the dS algebra. One might ask why, then, experimental data in particle physics give the impression that particles and antiparticles are fundamental.
As explained in Sec. \ref{subsecdS}, the matter is that, at this stage of the universe, the contraction parameter $R$ from the dS to Poincare algebra is very large and, in the formal limit $R\to\infty$,
one IR of the dS algebra splits into two IRs of the Poincare algebra corresponding to a particle and
its antiparticle with the same masses. In this case, for proving the equality of masses there is
no need to involve local covariant fields and the proof is given fully  
in the framework of (${\cal H,O,S}$).
As noted in Sec. \ref{DirEq}, in the spirit of the Dirac equation,
there should not be separate particles the electron and positron, but there should be one
particle combining them. In the case of dS symmetry, this idea
is implemented exactly in this way. It has been also noted that in this case
there are no conservation laws for additive quantum numbers: from the experiment it seems that such conservation laws take place, but in fact, these laws are only approximate because,
at the present stage of the universe the parameter $R$ is very large. {\it Thus, we can conclude that dS symmetry is more fundamental than Poincare and AdS symmetries.}

We now discuss the dS vs. AdS problem from the point of view whether standard gravity 
can be obtained in the framework of a free theory. In standard nonrelativistic approximation, 
gravity is characterized by the term $-Gm_1m_2/r$ in the mean value of the mass operator.
Here $m_1$ and $m_2$
are the particle masses and $r$ is the distance between
the particles. Since the kinetic energy is always positive,
the free nonrelativistic mass operator is positive definite
and therefore there is no way to obtain gravity in the
framework of a free theory. Analogously, in Poincare
invariant theory, the spectrum of the free two-body mass
operator belongs to the interval $[m_1+m_2,\infty )$ while the existence of gravity necessarily requires
that the spectrum should contain values less than $m_1+m_2$.

As explained in Sec. \ref{subsecAdS}, in theories where the symmetry algebra is the AdS algebra,   for positive
energy IRs, the AdS Hamiltonian has the spectrum in
the interval $[\mu,\infty )$ and $\mu >0$ has the meaning 
of the mass. Therefore the situation is pretty much
analogous to that in Poincare invariant theories.
In particular, the free two-body mass operator again
has the spectrum in the interval $[\mu_1+\mu_2,\infty )$
and therefore there is no way to reproduce gravitational
effects in the free AdS invariant theory.

In contrast to the situation in Poincare and AdS invariant theories, the free mass operator in dS theory
is not bounded below by the value of $\mu_1+\mu_2$. The discussion in Sec. \ref{subsecdS} shows that this property by no means implies that the theory is unphysical. In the dS case, there
is no law prohibiting that in the nonrelativistic approximation, the mean value of the mass operator contains the term $-Gm_1m_2/r$. Therefore if one has a choice between Poincare, AdS and dS symmetries then the only chance to describe gravity in a free theory is to choose dS symmetry, and, as discussed in \cite{book}, a possible nature of gravity
is that gravity is a kinematical effect in a quantum theory based not on complex numbers
but on a finite ring or field. {\it This is an additional argument in favor of dS vs. AdS.}

We now apply this conclusion to the known problem of baryon asymmetry of the universe
(BAU).
This problem is formulated as follows. 
According to modern particle and cosmological theories, the numbers of baryons and antibaryons in the early stages of the universe were the same. Then, since the baryon number is the conserved quantum number, those numbers should be the same at the present stage. However, at this stage, the number of baryons is much greater than the number of antibaryons.

However, as noted above, it seems to us that the baryon quantum number is conserved because at the present stage of the evolution of the universe, the value of $R$ is enormous. 
As noted in Sec.
\ref{symmetry}, it is reasonable to expect that $R$ changes over time, and as noted in Sec. \ref{explanation}, in semiclassical approximation, $R$ coincides with the radius of the universe. 
As noted in Sec. \ref{subsecdS}, even if $R$ is very large but finite then there is a non-zero
probability of transitions particle$\leftrightarrow$antiparticle.  But, according to cosmological theories, at early stages of the universe, $R$ was much
less that now. At such values of $R$, the very concepts of particles, antiparticles and baryon number do not have a physical meaning. {\it So, the statement that at early stages of the universe the numbers of baryons and antibaryons were the same, also does not have a physical meaning, and, as a consequence,  the BAU problem does not arise.}

\chapter{Solving cosmological constant problem}

\section{Introduction}
\label{intro}

At the present stage of the universe (when semiclassical approximation is valid), in the phenomenon of cosmological acceleration (PCA), only nonrelativistic macroscopic bodies are involved,
and one might think that here there is no need to
involve quantum theory. However, ideally, the results for every classical (i.e., non-quantum) problem should be obtained from quantum theory in semiclassical approximation.
We will see that, considering PCA from
the point of view of quantum theory sheds a new light on understanding this problem.

In PCA, it is assumed that the bodies are located at large (cosmological) distances from each other and sizes of the bodies are much less than distances between them. Therefore, interactions between the bodies can be neglected and, from the formal point of view, the description of our system
is the same as the description of $N$ free spinless elementary particles.

However, in the literature, PCA is usually
considered in the framework of dark energy and other exotic concepts. In Sec. \ref{Dark}
we argue that such considerations are not based on rigorous physical principles. In Sec.
\ref{symmetry} we have explained how symmetry should be defined at the quantum level,
and in Sec. \ref{explanation} we describe PCA in the framework of our approach.

\section{History of dark energy}
\label{Dark}

This history is well-known. Immediately after the creation of GR, Einstein believed that, since,
in his opinion, the universe is stationary, the cosmological constant $\Lambda$  in his equations must be non-zero, and this point of view has been described in his paper \cite{EinsteinLambda} written in 1917. On the other hand,
in 1922, Friedman found solutions of equations of GR with $\Lambda=0$ to provide theoretical evidence that the universe is expanding \cite{Friedman}. The author of \cite{Steer} states that Lundmark was the first person to find observational evidence for expansion in 1924 — three years before Lemaître and five years before Hubble, but, for some reasons, Lundmark’s research was not adopted and his paper was not published.
In 1927, Lemaître independently reached a similar conclusion to Friedman on a theoretical basis, and also presented observational evidence (based on the Doppler effect) for a linear relationship between distance to galaxies and their recessional velocity \cite{Lem}. In paper \cite{Hubble} written in 1929, Hubble described his results which observationally confirmed Lundmark's and Lemaître's findings.

According to Gamow's memories, after Hubble showed Einstein the results of observations at
the Mount Wilson observatory,   
Einstein said that introducing $\Lambda\neq 0$ was the biggest blunder of his life. 
After that, the statement that $\Lambda$ must be zero was advocated even in textbooks. 

The explanation was that, according to the philosophy of GR, matter creates a curvature of space-time, so when matter is absent, there should be no curvature, i.e.,
space-time background should be the flat Minkowski space. That is why when in 1998 it was realized that
the data on supernovae could be described only with $\Lambda\neq 0$, the impression was that it was a shock of something fundamental. However, the terms with $\Lambda$ in the Einstein equations have been moved from the left-hand side to the 
right-hand one, it was declared that in fact $\Lambda=0$, but the impression that $\Lambda\neq 0$ was the manifestation of a hypothetical field which, depending on the model, was called dark energy or quintessence. 
In spite of the fact that, as noted in wide publications (see e.g., \cite{Brax} and references therein), 
their physical nature remains a mystery, the most publications on PCA involve those concepts.

Several authors criticized this approach from the following considerations.
GR without the contribution of $\Lambda$ has been confirmed with a high accuracy in 
experiments in the Solar System. If $\Lambda$ is as small as it has been observed, it can
have a significant effect only at cosmological distances while for experiments in the
Solar System, the role of such a small value is negligible. The authors of \cite{Lambda}
titled ''Why All These Prejudices Against a Constant?'' note that it is not clear why
we should think that only a special case $\Lambda=0$ is allowed. If we accept the theory
containing the gravitational constant $G$ which is taken from outside, then why can't we accept a theory containing two independent constants?

Let us note that currently there is no physical theory which works under all conditions. For example, it is not correct to extrapolate nonrelativistic theory to cases when speeds are comparable to $c$ and to extrapolate classical physics for describing energy levels of the hydrogen atom. GR is a successful non-quantum theory for describing macroscopic phenomena where large masses are present, but extrapolation of GR to the case when matter disappears is not physical. One of the principles of physics is that a definition of a physical quantity is a description of how this quantity should be measured. As noted in Sec. \ref{PoincareIRs}, the concepts of space and its curvature are purely mathematical. Their aim is to describe the motion of real bodies. But the concepts of empty space and its curvature should not be used in physics because nothing can be measured in a space which exists only in our imagination.  
Indeed, in the limit of GR when matter disappears, space remains and has a curvature (zero curvature when 
$\Lambda=0$, positive curvature when $\Lambda>0$ and negative curvature when $\Lambda<0$) while, since space is only a mathematical concept for describing matter, a reasonable approach should be such that in this limit space should disappear too. 

A common principle of physics is that, when a new phenomenon is discovered, physicists should try to first explain it proceeding from the existing science. Only if all such efforts fail, something exotic can be involved.  But for PCA, an opposite approach was adopted: exotic explanations with dark energy or quintessence were accepted without serious efforts to explain the data in the framework of existing science.

Although the physical nature of dark energy and quintessence remains a mystery, there exists a wide literature where the 
authors propose QFT models of them. For example, as noted
in \cite{Kam1}, there are an almost endless number of explanations
for dark energy. While in most publications, only proposals about 
future discovery of dark energy are considered, the authors of \cite{Brax} argue
that dark energy has already been discovered by the XENON1T collaboration.
In June 2020, this collaboration reported an excess of electron recoils: 285 events, 53 more than expected 232 with a statistical significance of 
$3.5\sigma$. However, in July 2022, a new analysis by the XENONnT collaboration 
discarded the excess \cite{arxiv}.

Several authors (see e.g., \cite{Kam1,Kam2, Kam3}) proposed approaches where some
quantum fields manifest themselves as dark energy at early stages of the universe,
and some of them are active today. However, as shown in our publications and in the present paper, at least at the present stage of the universe (when semiclassical approximation is valid), PCA can be explained without uncertainties proceeding from universally recognized
results of physics and without involving models and/or assumptions the validity of which has not been unambiguously proved yet.

\section{Explanation of cosmological acceleration}
\label{explanation}

Standard particle theories involve self-adjoined IRs of the Poincare algebra. They are described even in textbooks and do not involve Minkowski space. Therefore, when Poincare symmetry is replaced by more general 
dS or AdS one, dS and AdS particle theories should be based on self-adjoined IRs of the dS or AdS algebras. However, physicists usually are not familiar with such IRs because they believe that dS and AdS quantum theories
necessarily involve quantum fields on dS or AdS spaces, respectively.

The mathematical literature on unitary IRs of the dS group is wide but there are only a few papers where such IRs are described for physicists. For example, the excellent Mensky's book \cite{Mensky} exists only in Russian.
At the same time, to the best of our knowledge, self-adjoint IRs of the dS algebra 
have been described from different considerations only in \cite{JPA,JMP,ECHAYA,symm}, and
some of those results have been mentioned in Sec. \ref{subsecdS}. It has been noted that
the space of an IR consists of functions defined on two hyperboloids
and in the limit $R\to\infty$ one IR of the dS algebra splits into two IRs of the Poincare
algebra with positive and negative energies.  

As noted in Sec. \ref{intro}, the results on IRs can be applied not only to elementary
particles but even to macroscopic bodies when it suffices to
consider their motion as a whole. We will consider this case and will
consider the operators $M^{4\mu}$ not
only in Poincare approximation but taking into account dS
corrections. If those corrections are small, it suffices to consider only states with
the support on the upper hyperboloid and describe the representation operators by Eq. (\ref{upperhyperb}).

We define the quantities $E,\,\,{\bf P},\,\,m$ by Eq. (\ref{EPm}) and consider the non-relativistic approximation when $|{\bf v}|\ll 1$. If we wish to work with units where the dimension of
velocity is $m/s$, we should replace ${\bf v}$ by ${\bf
v}/c$. If ${\bf p}=m{\bf v}$ then it is clear from the
expressions for ${\bf B}$ in Eq. (\ref{upperhyperb}) that ${\bf p}$ becomes the real 
momentum ${\bf P}$ only in the limit $R\to\infty$. 

The operators in Eq. (\ref{upperhyperb}) act in momentum representation and 
{\it at this stage, we have no spatial coordinates yet}. For describing the
motion of particles in terms of spatial coordinates, we must define the position operator.
A question: is there a law defining this operator? 
The postulate that the coordinate and momentum representations are related by the Fourier transform was taken at the dawn of quantum theory by analogy with classical electrodynamics, where the coordinate and wave vector representations are related by this transform. 
But the postulate has not been derived from anywhere, and there is no experimental confirmation of the postulate beyond the nonrelativistic semiclassical approximation. Heisenberg, Dirac, and others argued in favor of this postulate but, for example, in the problem of describing photons from distant stars, the connection between the coordinate and momentum representations should  be not through the Fourier transform, but as shown in \cite{book}. However,
since, PAC involves only nonrelativistic bodies then the position operator in momentum representation
can be defined as usual, i.e., as ${\bf r}=i\hbar \partial /\partial {\bf p}$, and in semiclassical approximation, we can treat ${\bf p}$ and ${\bf r}$ as usual vectors. 

Then as follows from Eq. (\ref{upperhyperb})
\begin{equation}
{\bf P}= {\bf p}+mc{\bf r}/R, \quad H = {\bf p}^2/2m +c{\bf p}{\bf r}/R,\quad {\bf N}=-m{\bf r}
\label{II64}
\end{equation}
where $H=E-mc^2$ is the classical nonrelativistic Hamiltonian.
As follows from these expressions
\begin{equation}
H({\bf P},{\bf r})=\frac{{\bf P}^2}{2m}-\frac{mc^2{\bf r}^2}{2R^2}
\label{II66}
\end{equation}
Here the last term is the dS correction to the non-relativistic Hamiltonian. 
Now it follows from the Hamilton equations 
that even one free particle is moving with the acceleration
\begin{equation}
{\bf a}={\bf r}c^2/R^2=\frac{1}{3}c^2\Lambda {\bf r}
\label{accel}
\end{equation}
where ${\bf a}$ is the acceleration, ${\bf r}$ is the
radius vector and $\Lambda=3/R^2$.  

The observed quantities are not absolute but relative with respect to a body that is chosen as the reference frame. We can take into account  that 
the representation describing a free N-body system is the tensor product of the corresponding single-particle IRs. It means that
every N-body operator $M^{ab}$ is a sum of the corresponding single-particle operators. 

Consider a system of two free particles described by the
variables ${\bf P}_j$ and ${\bf r}_j$ ($j=1,2$). Define 
standard nonrelativistic variables
\begin{eqnarray}
&&{\bf P}_{12}={\bf P}_1+{\bf P}_2,
\quad {\bf q}_{12}=(m_2{\bf P}_1-m_1{\bf P}_2)/(m_1+m_2)\nonumber\\
&&{\bf R}_{12}=(m_1{\bf r}_1+m_2{\bf r}_2)/(m_1+m_2),\quad
{\bf r}_{12}={\bf r}_1-{\bf r}_2
\label{II68}
\end{eqnarray}
Then explicit calculations (see e.g., Eq. (61) in \cite{JPA}, Eq. (17) in \cite{ECHAYA}
or Eq. (17) in \cite{symm}) give that the two-body mass is   
\begin{equation}
M({\bf q}_{12},{\bf r}_{12})=
m_1+m_2 +H_{nr}({\bf r}_{12},{\bf q}_{12}),\quad 
H_{nr}({\bf r},{\bf q})=\frac{{\bf q}^2}{2m_{12}}-\frac{m_{12}c^2{\bf r}^2}{2R^2}
\label{II70}
\end{equation}
where $H_{nr}$ is the internal two-body Hamiltonian and $m_{12}$ is the reduced two-particle mass. 
Then, as a consequence of the Hamilton equations, in semiclassical approximation, the relative
acceleration is again given by Eq. (\ref{accel}) but now
${\bf a}$ is the relative acceleration and ${\bf r}$ is the
relative radius vector. 

From a formal point of view, such a calculation must be carried out to confirm mathematically that here standard non-relativistic concepts work since, from the point of view of these concepts, if, for example, the acceleration of the first particle is ${\bf a}_1$ and of the second is ${\bf a}_2$, then their relative acceleration equals ${\bf a}_1-{\bf a}_2$. 

The fact that the relative acceleration of noninteracting bodies is not zero does not contradict the law of inertia, because this law is valid only in the case of Galilei and Poincare symmetries, and in the formal limit $R\to\infty$, ${\bf a}$ becomes zero as it should be.
Since $c$ is the contraction parameter for the transition from Poincare invariant theory to Galilei invariant one, the results of the latter can be obtained from the former in the formal limit $c\to\infty$, and Galilei invariant theories do not contain $c$. Then one might ask why Eq. (\ref{accel}) contains $c$ although we assume that
the bodies in PCA are nonrelativistic. The matter is that Poincare invariant theories do not contain $R$ but we work in dS invariant theory and assume that, although $c$ and $R$ are very large, they are not infinitely large, and the quantity $c^2/R^2$ in
Eq. (\ref{accel}) is finite.

As noted in Sec. \ref{asymm}, dS symmetry is more general than AdS one.
Formally, an analogous calculation using the results of Chap. 8 of
\cite{book} on IRs of the AdS algebra gives
that, in the AdS case, ${\bf a}=-{\bf r}c^2/R^2$, i.e., we have attraction instead of repulsion.  The
experimental facts that the bodies repel each other confirm that dS symmetry is indeed more
general than AdS one. 

The relative accelerations given by Eq. (\ref{accel}) are the same as those derived from GR if the curvature of dS space equals $\Lambda=3/R^2$, where $R$ is the radius of this space. 
{\it However, the crucial difference between our results and the results of GR is as follows.
While in GR, $R$ is the radius of the dS space and can be arbitrary, as explained in detail in Sec.
\ref{symmetry}, in quantum
theory, $R$ has nothing to do with the
radius of the dS space, it is the coefficient of proportionality between $M^{4\mu}$ and $P^{\mu}$, it is fundamental to the same extent as $c$ and $\hbar$, and a question why $R$ is as is does not arise.} {\bf Therefore, our approach gives a clear explanation why $\Lambda$ is
as is.}

The fact that two free particles have a relative acceleration
is known for cosmologists who consider dS symmetry at the
classical level. This effect is called the dS antigravity. The
term antigravity in this context means that particles
repulse rather than attract each other. In the case of the dS
antigravity, the relative acceleration of two free particles is
proportional (not inversely proportional!) to the distance
between them. This classical result (which in our approach has
been obtained without involving dS space and Riemannian
geometry) is a special case of dS symmetry at the quantum level
when semiclassical approximation works with a good accuracy.

As follows from Eq. (\ref{accel}), the dS antigravity is not important for local
physics when $r\ll R$. At the same time, at cosmological distances the dS antigravity is much stronger than any other
interaction (gravitational, electromagnetic {\it etc.}). One can consider the quantum two-body problem with the 
Hamiltonian $H_{nr}$ given by Eq. (\ref{II70}). Then it is obvious that the spectrum of the operator $H_{nr}$ is purely
continuous and belongs to the interval $(-\infty,\infty)$ (see also \cite{JPA,lev1b} for details).
This does not mean that the theory is unphysical since stationary bound states in standard theory
become quasistationary with a very large lifetime if $R$ is large.

In the literature it is often stated that quantum theory of gravity should become GR in classical approximation.
In Sec.\ref{DirEq} we argue that this is probably not the case because at the quantum level
the concept of space-time background does not have a physical meaning. {\it Our results for the cosmological acceleration obtained
from semiclassical approximation to quantum theory are compatible with GR} but in our approach, space-time background is absent from the very beginning.

In GR, the result (\ref{accel}) does not depend on how $\Lambda$ is interpreted,
as the curvature of empty space or as the manifestation of
dark energy. However, in quantum theory, there
is no freedom of interpretation. Here $R$ is the parameter of contraction from the dS Lie algebra
to the Poincare one, it has nothing to do with the radius of the background space and with dark energy and it  must be finite because dS symmetry is more general than Poincare one. 

\section{Discussion}
\label{discussion}

We have shown that, at the present stage of the universe (when semiclassical approximation
is valid), the phenomenon of cosmological acceleration is simply a {\it kinematical} consequence
of quantum theory in semiclassical approximation, and this conclusion has been made
without involving models and/or assumptions the validity of which has not been unambiguously proved yet. 

The concept of the cosmological constant $\Lambda$ has been originally defined in
GR which is the purely classical (i.e., not quantum) theory. Here $\Lambda$ is the curvature
of space-time background which, as noted in Secs. \ref{DirEq} and \ref{PoincareIRs},
is a purely classical concept. Our consideration does not
involve GR, and, as explained in Sec. \ref{symmetry}, the contraction parameter $R$ from dS invariant to Poincare invariant theory
has nothing to do with the radius of dS space. 

However, in QFT, $\Lambda$ is interpreted as vacuum energy density, and the 
cosmological constant problem is described in a wide literature (see e.g. \cite{CC}
and references therein). Usually, this problem is considered 
in the framework of Poincare invariant QFT of gravity on Minkowski space.
This theory contains only one phenomenological parameter --- the gravitational constant $G$,
and $\Lambda$ is defined by the vacuum expectation
value of the energy-momentum tensor. The theory contains strong divergencies
which cannot be eliminated because the theory is not renormalizable. The 
results can be made finite only with a choice of the cutoff parameter.
Since $G$ is the only parameter in the theory, the usual choice of the cutoff parameter
in momentum space is $\hbar/l_P$ where $l_P$ is the Plank length.
Then, if $\hbar=c=1$, $G$ has the dimension $length^2$ and $\Lambda$ is
of the order of $1/G$. This value is approximately by 122 orders of magnitude greater
than the experimental one, and this situation is called vacuum catastrophe. It is discussed in a wide literature how the discrepancy with experiment can be reduced, but the problem remains.

The approach to finding $\Lambda$ in terms of $G$ cannot be fundamental for several reasons. 
First of all, as noted in Sec. \ref{symmetry}, fundamental dS and AdS quantum theories originally do not
contain dimensional parameters. The dimensional quantities $(c,\hbar,R)$ can be introduced to
those theories only as contraction parameters for transions from more general theories to less
general ones. QFT of gravity is based on Poincare symmetry which is a special degenerate case of 
dS and AdS symmetries in
the formal limit $R\to\infty$. This theory contains $G$, but it is
not explained how $G$ is related to contraction from dS and AdS symmetries to Poincare symmetry.
Also, as noted in Sec. \ref{DirEq}, in quantum theories involving space-time background
the conditions (${\cal H,O,S}$) are not met and such theories contain mathematical
inconsistencies. The problem of constructing quantum theory of gravity is one of the most fundamental
problems in modern theory and the assumption that this theory will be Poincare invariant QFT is not based
on rigorous physical principles. 

{\bf In any case, as follows from the very problem statement about the cosmological acceleration, 
$\Lambda$ should not depend on $G$.} Indeed, as noted in Sec. \ref{intro}, in this problem,
it is assumed that the bodies are located at large (cosmological) distances from each other and sizes of the bodies are much less than distances between them. Therefore, all interactions between the bodies 
(including gravitational ones) can be neglected and, from the formal point of view, the description of our system
is the same as the description of $N$ free spinless elementary particles.

As explained in detail in Sec. \ref{symmetry}, at the quantum level, the parameter $R$ is fundamental to the same extent as $c$ and $\hbar$, it has nothing to do with the relation between Minkowski and
dS spaces and the problem why $R$ is as is does not arise by analogy with the problem
why $c$ and $\hbar$ are as are. As noted in Sec. \ref{explanation}, the results for cosmological acceleration in our approach 
 and in GR are given by the same expression (\ref{accel}) but the crucial difference between our approach and GR is as follows.
While in GR, $R$ is the radius of the dS space and can be arbitrary, in our approach, $R$ is defined uniquely because it is a parameter of contraction from the dS algebra to the
Poincare one. Therefore, our approach explains why the cosmological constant is as. 

{\it Therefore, at the present stage of the universe (when semiclassical approximation is valid), the phenomenon of cosmological acceleration has nothing to do with dark energy or other artificial reasons. This phenomenon is an inevitable {\bf kinematical} consequence of quantum theory in semiclassical approximation and the vacuum catastrophe and the problem of cosmological constant do not arise.}

Since 1998,  it has been confirmed in several experiments \cite{AA} that $\Lambda>0$, and 
$\Lambda = 1.3\cdot 10^{-52}/m^2$ with the accuracy 5\%. 
Therefore, {\it at the current stage of the universe}, $R$ is of the order of $10^{26}m$.
Since $\Lambda$ is very small and the evolution of the universe is the complex process, cosmological repulsion does not appear to be the main effect determining this process, and other effects
(e.g., gravity, microwave background and cosmological nucleosynthesis) may play a much larger role.

\chapter{Open problems}
\label{open}

As noted by Dyson in his fundamental paper \cite{Dyson}, nonrelativistic theory is a special
degenerate case of relativistic theory in the formal limit $c\to\infty$ and relativistic theory
is a special degenerate case of dS and AdS theories in the formal limit $R\to\infty$ and, as shown in Sec. \ref{asymm}, dS symmetry is more general than AdS one. 

The paper \cite{Dyson} appeared in 1972, i.e., more than 50 years ago, and, in view of Dyson's results, a question arises why general particle theories
(QED, electroweak theory and QCD) are still based on Poincare symmetry and not dS one. Probably physicists believe that, since, at least at the present stage of the universe,  $R$ is much greater than even sizes of stars, dS symmetry can play an important role only in cosmology and there is no need to use it for description of elementary particles.

We believe that this argument is not consistent because usually more general theories shed a new light on standard concepts. It is clear from the discussion in Sec. \ref{asymm} that the 
construction of dS theory will be based on considerably new concepts than the
construction of standard quantum theory because in dS theory the concepts
of particles, antiparticles and additive quantum numbers (electric charge, baryon quantum number and others) can be only approximate.

Another problem discussed in a wide literature is that supersymmetric generalization exists
in the AdS case but does not exist in the dS one. It may be a reason why supersymmetry
has not been discovered yet.

In \cite{book} we have proposed a criterion when theory A is more general (fundamental)
than theory B:

{\bf Definition: }{\it Let theory A contain a finite nonzero parameter and theory B be obtained from theory A in the formal limit when the parameter goes to zero or infinity. Suppose that with any desired accuracy theory A can reproduce any result of theory B by choosing a value of the parameter. On the contrary, when the limit is already
taken then one cannot return back to theory A and theory B cannot reproduce all results of theory A. Then theory A is more general than theory B and theory B is a special degenerate case of theory A}. 

As shown in \cite{book}, by using this {\bf Definition} one can prove that: a) nonrelativistic theory is a special degenerate case of relativistic theory
in the formal limit $c\to\infty$; b) classical theory is a special degenerate case of quantum theory
in the formal limit $\hbar\to 0$; c) relativistic theory is a special degenerate case of 
dS and AdS theories in the formal limit $R\to\infty$; d) standard quantum theory (SQT) based on complex numbers is a special degenerate case of finite quantum theory (FQT) based on finite mathematics  with a ring or field of characteristic $p$ in the formal limit $p\to\infty$. 

As noted in Sec. \ref{symmetry}, the properties a)-c) take place in SQT, and below we will
discuss the property d). As described in Secs. \ref{subsecAdS} and \ref{subsecdS},
in IRs of the AdS algebra, the energy spectrum of the energy operator can be either positive or negative while in the dS case, the spectrum necessarily contains energies of both signs.
As explained in Sec. \ref{asymm}, for this reason, the dS case is more physical than the
AdS one. We now explain that in the FQT analog of the AdS symmetry the situation is analogous to that in the dS case of SQT. For definiteness, we consider the case when $p$ is odd. 

By analogy with the construction of positive energy IRs in SQT, in FQT we start the construction from the rest state, where the AdS energy is positive and equals $\mu$. Then we act  on this state by raising operators and gradually get states with higher and higher energies, i.e., $\mu+1,\mu+2,...$. However, in contrast to the situation in SQT, we cannot obtain infinitely large numbers. When we reach the state with the energy $(p-1)/2$, the next state has
the energy $(p-1)/2+1=(p+1)/2$ and, since the operations are modulo $p$, this value also can be denoted as $-(p-1)/2$ i.e., it may be called negative.
When this procedure is continued, one gets the energies $-(p-1)/2+1=-(p-3)/2,-(p-3)/2+1=-(p-5)/2,...$
and, as shown in \cite{book}, the procedure finishes when the energy $-\mu$ is reached.

Therefore the spectrum of energies contains the values $(\mu,\mu+1,...,(p-1)/2)$ and 
$(-\mu,-(\mu+1),...,-(p-1)/2)$ and in the formal limit $p\to\infty$, this IR splits into two IRs of the AdS algebra in SQT for a particle  
with the energies $\mu,\mu+1,\mu+2, ...\infty$ and antiparticle with the energies
$-\mu,-(\mu+1),-(\mu+2), ...-\infty$  and both of them have the same mass $\mu$.
We conclude that in FQT, all IRs necessarily contain states with both, positive and
negative energies and the mass of a particle automatically equals the mass of the
corresponding antiparticle. This is an example when FQT can solve a problem which standard quantum AdS theory
cannot. By analogy with the situation in the standard dS case, for combining a particle and its antiparticle together, there is no need to involve additional coordinate fields because a particle and its antiparticle are already combined in the same IR.

Since the AdS case in FQT satisfies all necessary physical conditions, it is reasonable to investigate whether this case has a supersymmetric generalization. We first note that
representations of the standard Poincare superalgebra are described by
14 operators. Ten of them are the representation
operators of the Poincare algebra---four momentum operators and
six operators of the Lorentz algebra, and in addition,
there are four fermionic operators. The anticommutators
of the fermionic operators are linear combinations of the Lorentz algebra operators, the commutators of the fermionic
operators with the Lorentz algebra operators are linear
combinations of the fermionic operators and the
fermionic operators commute with the momentum operators. However,
the latter are not bilinear combinations of fermionic operators.

From the formal point of view, representations of the AdS superalgebra osp(1,4)
are also described by 14 operators --- ten
representation operators of the so(2,3) algebra and four
fermionic operators. There are three types of relations: the
operators of the so(2,3) algebra commute with each other as in Eqs. (\ref{CR}),
anticommutators of the fermionic
operators are linear combinations of the so(2,3) operators and
commutators of the latter with the fermionic operators are
their linear combinations. However, representations of
the osp(1,4) superalgebra can be described exclusively in terms
of the fermionic operators. The matter is that anticommutators of four operators form ten
independent linear combinations. Therefore, ten bosonic
operators can be expressed in terms of fermionic ones. This is
not the case for the Poincare superalgebra since it is obtained from the so(2,3) one by
contraction. One can say that the representations of the
osp(1,4) superalgebra is an implementation of the idea that
supersymmetry is the extraction of the square root from the
usual symmetry (by analogy with the treatment of the
Dirac equation as a square root from the Klein-Gordon one).
From the point of view of the osp(1,4) supersymmetry, only four fermionic operators are
fundamental, in contrast to the case when in dS and AdS symmetries there are
ten fundamental operators.

As noted in Sec. \ref{subsecdS}, in the approach when a particle and its antiparticle belong
to the same IR, it is not possible to define the concept of neutral particles. For example, a problem arises whether the photon is the elementary particle. In Standard Model
(based on Poincare invariance) only massless particles are treated as elementary.
However, as shown in the seminal paper by Flato and Fronsdal \cite{FF}
(see also \cite{HeidenreichS}), in standard AdS theory, each massless IR can be 
constructed from the tensor product of two singleton IRs and, as noted in \cite{book},
this property takes place also in FQT. The concept of singletons has been proposed
by Dirac in his paper \cite{DiracS} titled ''A Remarkable Representation of the 3 + 2 de Sitter group'', and, as discussed in \cite{book}, in FQT this concept is even more remarkable
than in SQT. As noted in Sec. \ref{subsecAdS}, even the fact that the AdS mass of the electron
is of the order of $10^{39}$ poses a problem whether the known elementary particles
are indeed elementary. In \cite{book} we discussed a possibility that only Dirac
singletons are true elementary particles.

As explained in \cite{book}, in FQT, physical quantities can be only finite, divergences cannot exist in principle,
and the concepts of particles, antiparticles, probability and additive quantum numbers can
be only approximate if $p$ is very large. 
The construction of FQT is one of the most fundamental (if not the most fundamental) problems of quantum theory. 

The above discussion indicates that fundamental quantum theory has a very long
way ahead (in agreement with Weinberg's opinion \cite{Wein2} that a new theory may be
centuries away).

{\bf Acknowledgments}. The author is grateful to AE, SIE, Vladimir Karmanov, Dmitry Kazakov, Maxim Khlopov, Ethan Vishnyak and the reviewers of this paper for useful remarks.

\begin{flushleft}{\bf Funding:} This research received no external funding.\end{flushleft}
\begin{flushleft}{\bf Data Availability Statement:} Not applicable. \end{flushleft}
\begin{flushleft}{\bf Conflicts of Interest:} The author declares that the research was conducted in the absence of any commercial or financial relationships that could be construed as a potential conflict of interest. \end{flushleft}

\end{document}